\documentclass[11pt,a4paper]{article}
\pdfoutput=1
\usepackage{jheppub}
\usepackage{subfig}

\newcommand{\ba}{\begin{eqnarray}}
\newcommand{\ea}{\end{eqnarray}}
\newcommand{\Tr}{\textrm{Tr}}
\newcommand{\barray}{\begin{array}}
\newcommand{\earray}{\end{array}}

\title{Cold Baryogenesis from first principles in the Two-Higgs Doublet model with Fermions}

\author[a]{Zong-Gang Mou,}
\author[a]{Paul M. Saffin,}
\author[b]{Anders Tranberg}

\affiliation[a]{School of Physics and Astronomy, University Park, University of Nottingham,\\ Nottingham NG7 2RD, United Kingdom}
\affiliation[b]{Faculty of Science and Technology, University of Stavanger, \\4036 Stavanger, Norway}

\emailAdd{ppxzm1@nottingham.ac.uk}
\emailAdd{paul.saffin@nottingham.ac.uk}
\emailAdd{anders.tranberg@uis.no}

\abstract{We present a first-principles numerical computation of the baryon asymmetry in electroweak-scale baryogenesis. For the scenario of Cold Baryogenesis, we consider a one fermion-family reduced CP-violating two Higgs-doublet model, including a classical SU(2)-gauge/two-Higgs sector coupled to one quantum left-handed fermion doublet and two right-handed singlets. Separately, the C(CP) breaking of the two-Higgs potential and the C and P breaking of the gauge-fermion interactions do not provide a baryon asymmetry. Only when combined does baryogenesis occur. Through large-scale computer simulations, we compute the asymmetry for one particularly favourable scalar potential. The numerical signal is at the boundary of what is numerically discernible with the available computer resources, but we tentatively find an asymmetry of $|\eta|\leq 3.5\times 10^{-7}$. }


\begin{document}

\maketitle

\section{Introduction}
\label{sec:intro}

Electroweak baryogenesis is still the subject of significant scientific attention, being a very elegant and testable set of scenarios to explain the baryon asymmetry of the Universe. The central element is the non-perturbative violation of baryon and lepton number in the electroweak sector of the Standard Model, which when combined with C- and CP-breaking interactions out of equilibrium allows for baryogenesis \cite{kuzmin}. Since the Standard Model and its simplest extensions (multiple scalar fields, massive leptons) break C and CP, there only remains to compute the asymmetry. This turns out to be very challenging to do from first principles, since it involves non-perturbative quantum dynamics of in particular fermions out of thermal equilibrium, and because the final asymmetry one is trying to compute is very small \cite{planck}
\ba
\eta=\frac{n_B}{n_\gamma}\simeq (6.0\pm 0.1)\times 10^{-10}.
\ea
Traditionally, the approach has been to split the computation up into a number of equilibrium and out-of-equilibrium quantities, which may each be treated using dedicated techniques. The state-of-the-art is quite advanced, and often involves considering effectively bosonic systems obtained through integrating out the fermions and/or dimensional reduction \cite{kajantie2}.  This has allowed a quite precise determination of the sphaleron rate \cite{donofrio}; the phase diagram of the Standard Model and its extensions \cite{kajantie,damgaard}; the bubble nucleation rate (in case of a first-order transition)\cite{mooreb}; the dynamics of bubbles and their observational signatures \cite{huber1,huber2}; the effective CP-violation in the Standard Model \cite{brauner,brauner2}; the interaction of a bubble wall with fermions (see for instance \cite{schmidt1,schmidt2,schmidt3}). Putting these quantities together gives a handle on the baryon asymmetry produced, and some are even no-go results, such as the non-existence of a first-order phase transition and the very strong suppression of Standard Model CP-violation. Historically, these no-go results removed the need for actually calculating a {\it number} for the asymmetry, apart from order-of magnitude estimates based on maximally favourable parameters. 

The need for a strong phase transition is in the Cold Baryogenesis scenario replaced by the requirement of a cold tachyonic transition at the end of inflation \cite{krauss,garciabellido,copeland}. A compelling feature of this scenario is the option of numerically computing the final asymmetry from first principles, simply because the mechanism is very simple: the particle creation process of reheating during an electroweak spinodal transition is asymmetric in particles and antiparticles. Such simulations have been performed for more than a decade \cite{turok,rajantie,garciabellido2,smit1,smit2,smit3,tranberg}, showing that the scenario is viable, given a cold initial condition. The simulations were however done in purely bosonic versions of the Standard Model or extensions with a singlet \cite{diaz} or a doublet scalar field \cite{wu1}. 

For electroweak baryogenesis, it is crucial to note that C-, CP- but also P-breaking is necessary to generate a baryon asymmetry. This follows from the anomaly equation relating the Chern-Simon number of the SU(2) gauge field and the baryon and lepton numbers
\ba
B(t)-B(0) = n_f\left[N_{\rm cs}(t)-N_{\rm cs}(0)\right]=L(t)-L(0),
\ea
and the observation that Chern-Simons number is odd under P but even under C. What this means in practice is that it is not enough to break C, thereby breaking CP if P is conserved. And it is also not enough to break P and thereby break CP if C is conserved. One needs to break C {\it and} P {\it and} CP, and this is indeed achieved in the Standard Model through the left-handed coupling to gauge fields (C and P broken maximally, CP conserved), and the complex phase in the CKM matrix (CP broken, by breaking C). Alternatively, in the two-Higgs doublet model (ignoring the CKM matrix), the left-handed coupling to gauge fields again provides C and P breaking, and in addition, complex parameters in the Higgs-Higgs potential break C and thereby CP. And only when combining them is an asymmetry produced. 

The results of different combinations of the breaking of discrete symmetries was demonstrated in \cite{smit1} for the bosonic part of the Standard Model with an effective CP and P-violating operator. One finds non-zero $N_{\rm cs}$, and the fermion number was inferred through the anomaly equation. The subsequent task is then to compute the coefficient of such an operator from integrating out the fermions in the full theory, including both C, P and CP-breaking. This turns out to be a substantial calculation, yielding a slightly different set of effective operators \cite{salcedo,smit,brauner}. 

In the bosonic part of the two-Higgs doublet model, it was seen that simply adding the C-breaking potential does not produce a net Chern-Simons number; only when in addition including an effective C- and P-breaking, but CP-conserving operator is an asymmetry produced. In \cite{wu2}, the operator
\ba
\delta V = \frac{\delta_{C/P}}{16\pi^2 m_W^2}i(\phi_1^\dagger\phi_2-\phi_2^\dagger\phi_2)\textrm{Tr}F^{\mu\nu}\tilde{F}_{\mu\nu},
\ea
was considered.
Again, it remains to compute the coefficient $\delta_{C/P}$ of this operator, this time from the CP-even fermion sector\footnote{If one chooses to neglect the contribution from CKM matrix.}. This has not been attempted beyond the simplest estimates \cite{turok}, but may in principle be done using the methods of \cite{salcedo,brauner}.

The obvious, but as it turns out far from straightforward, alternative is to include the fermions in the real-time dynamics. Fermions are inherently quantum mechanical, but may be formulated on a classical bosonic background \cite{aarts}. The numerical solution is computationally extremely demanding, but possible using what is known as "ensemble" fermions \cite{borsanyi}. Crucially, the anomaly equation carries through and the fermion back reaction on the bosons is reliably described \cite{saffin,mou}. The C- and P-breaking of the gauge-fermion coupling may now be directly introduced, and in combination with the C-breaking of the two-Higgs potential, an asymmetry should be created. The questions we wish to address are then the following:
\begin{itemize}
\item Is an asymmetry created in a cold spinodal transition?
\item Is this asymmetry large enough that we can see it numerically on the lattice?
\item Is it comparable to or ideally larger than the observed asymmetry?
\item Can we connect these results to previous work \cite{wu2}, for instance to provide an estimate of the coefficient $\delta_{C/P}$?
\item What is the numerical effort involved, and is it realistic to sweep a multidimensional experimentally allowed parameter space? 
\end{itemize}
Many of these questions will depend on the speed of the spinodal transition, the numerical effort available and the parameter range adopted for the Higgs-Higgs potential. But for moderately favourable choices, the answers are: yes, perhaps, yes, yes and unlikely.

In Section \ref{sec:model} we will introduce the reduced Standard Model that we will consider, including only one generation of fermions. We will set out the observables, numerical implementation, initial condition and the Higgs potential. In Section \ref{sec:results} we explain how to compute the baryon asymmetry and describe our numerical results. We conclude in Section \ref{sec:conclusion}. A number of technical details are relegated to a set of Appendices.

\section{The reduced Standard Model}
\label{sec:model}

We will consider a reduced version of the Two-Higgs Doublet Standard Model, where the SU(3) and U(1) interactions are ignored, and we only keep one generation of fermions, including a left-handed quark SU(2) doublet, $q_L =(u_L,d_L)^T$, a left-handed lepton SU(2) doublet, $l_L=(e_L,\nu_L)^T$ and right-handed singlets $u_R$, $d_R$ and $\nu_R$, $e_R$. Although we use a notation suggesting the first generation of the SM $(u,d,e,\nu_e)$, we expect the main effect of CP-violation to come from the heaviest fermions, in practice the third generation $(t,b,\tau,\nu_\tau)$. We emphasise that although for SM CP-violation it is crucial to have 3 generations of fermions (allowing for a complex phase in the CKM matrix to be physical), for the 2HDM CP-violation considered here, this is not necessary. Adding up to three families is straightforward, but require three times as much computational time.

\subsection{Continuum action}

Having made these simplifications, the continuum action reads
\ba
S&=&-\int\;d^4x\;\bigg[\frac{1}{4g^2}A^a_{\mu\nu}A^{a,\mu\nu} +
(D_\mu \phi_1)^\dagger (D^\mu \phi_1)
+(D_\mu \phi_2)^\dagger (D^\mu \phi_2)
\nonumber\\& &~~~\qquad\qquad+\bar{q}_L\gamma^\mu D_\mu q_L  +\bar{u}_R\gamma^\mu \partial_\mu u_R+\bar{d}_R\gamma^\mu \partial_\mu d_R
\nonumber\\& &~~~\qquad\qquad+\bar{l}_L\gamma^\mu D_\mu l_L +\bar{\nu}_R\gamma^\mu \partial_\mu \nu_R+\bar{e}_R\gamma^\mu \partial_\mu e_R
\nonumber\\& &~~~\qquad\qquad+V(\phi_1, \phi_2) + Y(q, l, \phi_1, \phi_2)
\bigg],
\ea
where the SU(2) gauge covariant derivatives are
\ba
D_\mu \phi = (\partial_\mu - i A_\mu) \phi,
\qquad                                                        
D_\mu  q_L = (\partial_\mu - iA_\mu) q_L,
\qquad                                                        
D_\mu  l_L = (\partial_\mu -iA_\mu) l_L,
\ea
and we have ordinary partial derivatives for the right-handed singlet fermion fields $u_R$, $e_R$, $d_R$ and $\nu_R$.

\subsubsection{Higgs potential}
\label{eq:potential}

\begin{figure}
\begin{center}
\includegraphics[width=10cm]{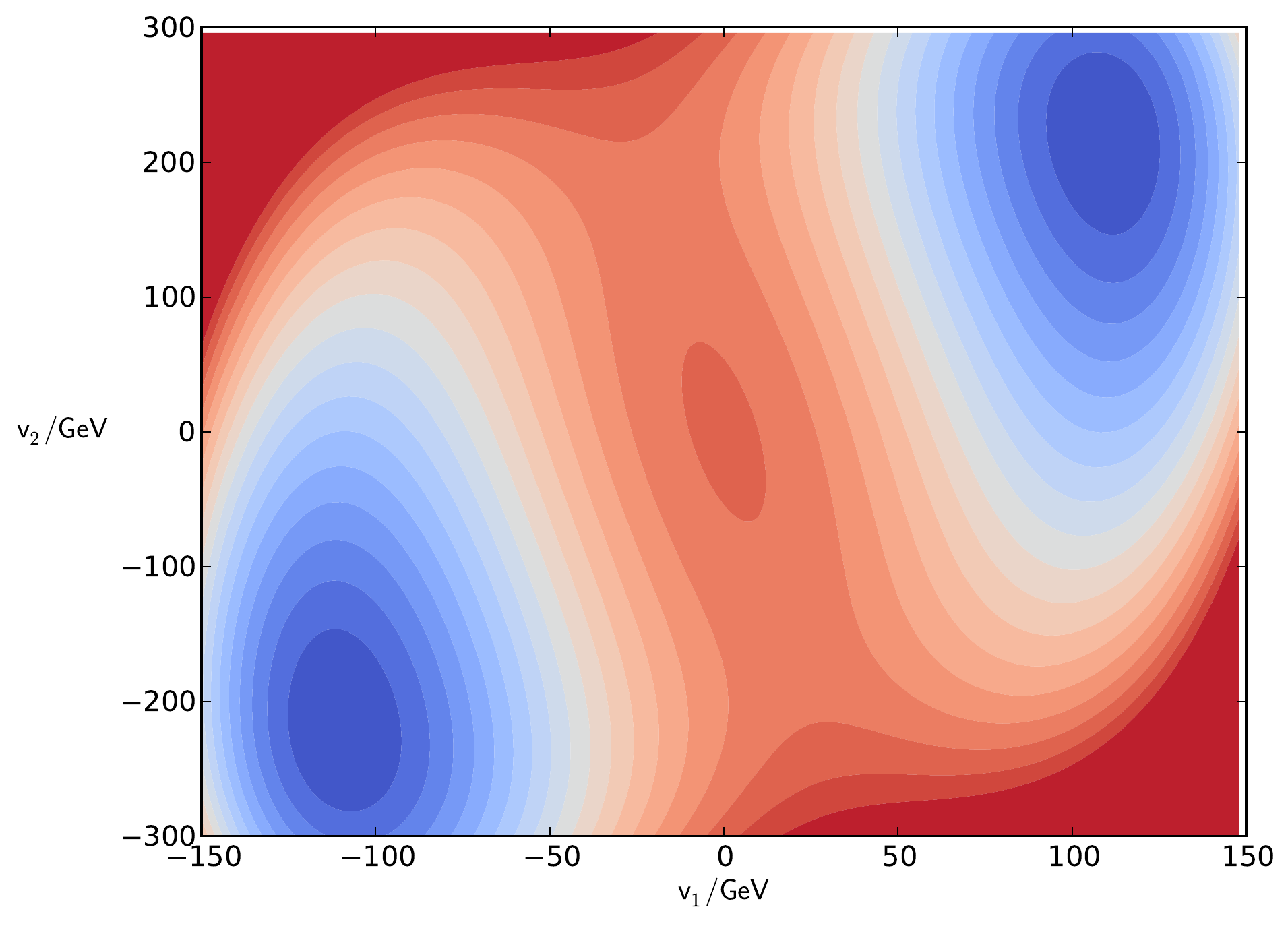}
\caption{The Higgs potential as a function of the unitary gauge Higgs field components $|v_1|$, $|v_2|$. The two degenerate minima are at $(v_1,v_2)=(\pm110, \pm220)$ GeV.}
\label{fig:potential}
\end{center}
\end{figure}

The two-Higgs scalar potential is given by
\ba
V(\phi_1, \phi_2)&=& 
-\frac{\mu_{11}^2}{2} (\phi_1^\dagger\phi_1)
-\frac{\mu_{22}^2}{2} (\phi_2^\dagger\phi_2)
-\frac{\mu_{12}^2}{2} (\phi_1^\dagger\phi_2)
-\frac{\mu_{12}^{2,\star} }{2}(\phi_2^\dagger\phi_1)
\nonumber\\&~&
+\frac{\lambda_1}{2} (\phi_1^\dagger \phi_1)^2 + \frac{\lambda_2}{2} (\phi_2^\dagger \phi_2)^2 + \lambda_3 (\phi_1^\dagger \phi_1) (\phi_2^\dagger \phi_2) + \lambda_4 (\phi_2^\dagger \phi_1) (\phi_1^\dagger \phi_2)
\nonumber\\&~&
+\frac{\lambda_5}{2} (\phi_1^\dagger \phi_2)^2 + \frac{\lambda_5^\star}{2} (\phi_2^\dagger \phi_1)^2.
\ea
This is the standard parametrisation \cite{PDG}, where we have chosen to put the couplings $\lambda_{6,7}$ to zero for simplicity. Allowing $\lambda_5$ and/or $\mu_{12}^2$ to be complex provides for C-violation at tree-level. If in addition $\textrm{Im}(\mu_{12}^2)=2 |v_1| |v_2|\textrm{Im}(\lambda_5)$, both Higgs vevs $v_{1,2}$ can be chosen real. Otherwise the vevs will have a relative phase $\textrm{Arg}(v^\dagger_2 v_1) $. We will choose this argument to be maximal, 
\ba
\textrm{Arg}(v^\dagger_2 v_1) &=& \frac{\pi}{2}.
\ea
Not all sets of complex $\lambda_5$ and $\mu_{12}$ lead to C-violation. A useful parametrisation is provided by \cite{osland}, which guides our choices below. We wish to maximise the effects of CP-violation, by tuning the parameters of the Higgs potential (the C- and P- breaking in the gauge-fermion coupling is already maximal). In addition we wish to have well separated Higgs mode masses, and we choose
for the neutral Higgs bosons
\ba
m_1=125\textrm{ GeV},\quad
m_2=300\textrm{ GeV},\quad
m_3=350\textrm{ GeV},
\ea
and for the charged modes
\ba
m_{\pm}=400\textrm{ GeV}.
\ea
We have also fixed the total Higgs vev, so that
\ba
|v_1|=110\textrm{ GeV},\qquad |v_2|=220\textrm{ GeV},\qquad \tan\beta=2,
\ea
with a relative phase of $\pi/2$ and 
\ba
|v_1|^2+|v_2|^2=(246\textrm{ GeV})^2.
\ea
After a number of tests at small lattice volumes, we settled on 
\ba
\mu_{11}^2  =              (233\textrm{ GeV})^2,\qquad
\mu_{22}^2  =                (311  \textrm{ GeV})^2,\qquad
\mu_{12}^2  =               (78.6  \textrm{ GeV})^2+ i(200 \textrm{ GeV})^2,\nonumber\\
\ea
with
\ba
\lambda_1 =                      5.2,\qquad
\lambda_2 =                      2.0,\qquad
\lambda_3 =                      5.1,\qquad
\lambda_4 =                      -4.2,\qquad
\lambda_5 =                    -0.56 + i0.26,
\ea
which amount to the angles mixing angles of the three neutral Higgs modes \cite{osland}
\ba
(\alpha_1, \alpha_2, \alpha_3) = (0.314, -1.49, 2.58),
\ea
The potential breaks C, due to the non-zero values of
\ba
\textrm{Arg}(\mu_{12}^2) = 0.45 \pi,\qquad
\textrm{Arg}(\lambda_5) = -0.14 \pi.
\ea
In Figure \ref{fig:potential}, we show a contour plot of the potential in the $|v_1|$, $|v_2|$ plane. The potential energy in the symmetric phase is
\ba
V_0=V(\phi_1=\phi_2=0)-V(v_1,v_2)=(226\textrm{ GeV})^4.
\ea

\subsubsection{Yukawa couplings}

The Yukawa coupling terms can in all generality be parametrized as
\ba
Y(q, l, \phi_1, \phi_2)&=&
~~ 
 G_1^d\bar{q}_L\phi_1 d_R
+G_2^d\bar{q}_L\phi_2 d_R
+G_1^e\bar{l}_L\phi_1 e_R
+G_2^e\bar{l}_L\phi_2 e_R
\nonumber\\&~&
+G_1^u\bar{q}_L\tilde\phi_1 u_R
+G_2^u\bar{q}_L\tilde\phi_2 u_R
+G_1^\nu\bar{l}_L\tilde\phi_1\nu_R
+G_2^\nu\bar{l}_L\tilde\phi_2\nu_R
\nonumber\\&~&
+G_1^{d\dagger}\bar  d_R \phi_1^\dagger q_L
+G_2^{d\dagger}\bar  d_R \phi_2^\dagger q_L
+G_1^{e\dagger}\bar  e_R \phi_1^\dagger l_L
+G_2^{e\dagger}\bar  e_R \phi_2^\dagger l_L
\nonumber\\&~&
+G_1^{u\dagger}  \bar      u_R      \tilde\phi_1^\dagger  q_L
+G_2^{u\dagger}  \bar      u_R      \tilde\phi_2^\dagger  q_L
+G_1^{\nu\dagger}\bar      \nu_R    \tilde\phi_1^\dagger  l_L
+G_2^{\nu\dagger}\bar      \nu_R    \tilde\phi_2^\dagger  l_L.
\ea
For simplicity, we will use a single coupling constant, 
\ba
\lambda_{\rm yuk}=G_2^d=G_2^e=G_1^u=G_1^\nu, \qquad G_1^d=G_1^e=G_2^u=G_2^\nu=0,
\ea
so that
\ba
Y(q, l, \phi_1, \phi_2)&=& 
~~ y_{\rm yuk}(\bar{q}_L\phi_2 d_R+\bar{l}_L\phi_2 e_R
+ \bar{q}_L\tilde\phi_1 u_R+ \bar{l}_L\tilde\phi_1\nu_R
\nonumber\\&~&
~ ~~~~+\bar d_R\phi_2^\dagger q_L+\bar e_R\phi_2^\dagger l_L
+ \bar u_R\tilde\phi_1^\dagger q_L + \bar\nu_R\tilde\phi_1^\dagger l_L).
\ea
There is no issue in principle using general Yukawa couplings, although it may become a little cumbersome. We take 
\ba
\lambda_{\rm yuk}=0.1.
\ea
It was demonstrated in \cite{mou} that the ensemble averaging procedure is under control for such large Yukawa couplings, which correspond to fermion masses of about $11/\sqrt{2}\simeq 8$ and $22/\sqrt{2}\simeq16$ GeV, depending on which Higgs field they are coupled to, easily larger than all the quark masses except for the top mass. 

\subsection{Asymmetric observables}
\label{sec:obs}

The baryon and lepton numbers are the spatial integrals over the zero-component of the baryon and lepton currents
\ba
j^\mu_{(b)}&=&i\left[\bar q_L\gamma^\mu q_L+\bar u_R\gamma^\mu     u_R+\bar d_R\gamma^\mu d_R\right]=i\bar q\gamma^\mu q,\\
j^\mu_{(l)}&=&i\left[\bar l_L\gamma^\mu l_L+\bar \nu_R\gamma^\mu \nu_R+\bar e_R\gamma^\mu e_R\right]=i\bar l\gamma^\mu l.
\ea
We will assign baryon number 1 to the quarks to emulate summing over SU(3) colour. And lepton number 1 to the lepton field.
Fermions coupled chirally to an SU(2) gauge field will experience a quantum anomaly, so that these currents obey
\ba
\partial_\mu j^\mu_{(b)}&=&\partial_\mu j^\mu_{(l)}=\frac{1}{32\pi^2}\left[\frac{1}{2}\epsilon^{\mu\nu\rho\sigma}F^a_{\mu\nu}F^a_{\rho\sigma}\right],\\
   &=&\partial_\mu K^\mu,
\ea
for each fermion doublet. The Chern-Simons current is given by 
\ba
K^\mu&=&\frac{1}{16\pi^2}\epsilon^{\mu\nu\rho\sigma}\left[F^a_{\nu\rho}A^a_\sigma-\frac{2}{3}\epsilon_{abc}A^a_\nu A^b_\rho A^c_\sigma\right],
\ea
and we therefore have that under a change of Chern-Simons number $N_{\rm CS}$ over time, we pick up a change in baryon and lepton number of
\ba
B(t)-B(0)=L(t)-L(0)=n_f[N_{\rm cs}(t)-N_{\rm cs}(0)]=\int_0^t dt \int d^3x \frac{1}{32\pi^2}\left[\frac{1}{2}\epsilon^{\mu\nu\rho\sigma}F^a_{\mu\nu}F^a_{\rho\sigma}\right].\nonumber\\
\ea
The Chern-Simons number changes continuously between one gauge vacuum and the next, and is an integer in the infinite series of (almost) degenerate gauge vacua. In contrast, the Higgs field winding number
\ba
N_W&=&\frac{1}{24\pi^2}\int d^3x\epsilon_{ijk} 
\hat \phi^\dagger\partial_i \hat \phi
\hat \phi^\dagger\partial_j \hat \phi
\hat \phi^\dagger\partial_k \hat \phi
,\qquad\qquad 
\hat \phi=\frac{1}{|\phi|} \phi,
\ea
is always an integer, changes discontinuously halfway between vacua, and coincides with Chern-Simons number in the vacua. Because it is an integer, the winding number turns out to be a better observable to count baryons, since if at the end of a simulation one has a well-defined $N_W$, one knows that because the sphaleron barriers are now up, $N_{\rm cs}$ and therefore $B$ and $L$ will eventually relax to the same integer value. 
We also note that there are two winding numbers $N_W^{1,2}$, one for each Higgs field, and they both coincide with the Chern-Simons number in the vacuum. Because of lattice artefacts, these are in fact not numerically exactly integers, and the jumps not strictly discontinuous. In addition to the expectation of the Higgs fields and the various energy components, our prime observables will be Chern-Simons number, the two winding numbers and the fermion number of the quark and lepton fields. 

\subsection{CP-symmetric initial conditions}
\label{sec:initconds}

We will consider an instantaneous quench of the Higgs potential at zero temperature. This in practice means that we evolve the initial configurations using the potential (\ref{eq:potential}) for times $t>0$ but that we generate these initial conditions in the vacuum corresponding to the quadratic potential
\ba
V(t<0) = \frac{\mu_{11}^2}{2} (\phi_1^\dagger\phi_1)+\frac{\mu_{22}^2}{2} (\phi_2^\dagger\phi_2)
+\frac{\mu_{12}^2}{2} (\phi_1^\dagger\phi_2)
+\frac{\mu_{12}^{2,\star} }{2}(\phi_2^\dagger\phi_1).
\ea
The vacuum is defined through the "half" method of \cite{rajantie}, which is to have field correlators and momentum correlators obey
\ba
\langle\phi^i_{\bf k}(\phi^j_{\bf k})^\dagger\rangle = \frac{1}{2\omega_{\bf k}}\delta_{ij},\qquad \langle\pi^i_{\bf k}(\pi^j_{\bf k})^\dagger\rangle = \frac{\omega_{\bf k}}{2}\delta_{ij},
\ea
where $\pi^i_{\bf k}$ and $\phi^i_{\bf k}$ are the momentum space variables for each real component of the Higgs fields (8 in total), and $\omega_{\bf k}^2=M^2+k^2$, with $M^2$ the eigenvalues of the mass matrix. The understanding being, that if $m_{12}$ is non-zero, one should first diagonalise in field space, generate the configuration in this basis and then rotate back to the $\phi_{1,2}$ basis. This prescription is the natural analog of the quench in \cite{smit3}, where in a single-Higgs model, the coefficient of the quadratic term "flips" instantaneously.

For each randomly generated configuration $\phi^i$, $\pi^i$, one also generates an entire ensemble of $N_f=4000$ fermion configurations ($u$, $d$, $e$, $\nu$ separately, with "male" and "female" copies \cite{borsanyi}). Since the initial state is the symmetric vacuum ($\langle\phi_1\rangle=\langle\phi_2\rangle=0$), the fermions are taken to be massless initially. Setting $A_\mu=0$ initially, we solve Gauss law for the SU(2) electric field $E_i$ in the background of the scalar-fermion fields. 

The behaviour of each of our fields under CP-transformations is
\ba
q_L^{cp}(t,x)& =i\gamma^0\gamma^2 ~ q^\star_L(t,-x) ,\qquad\qquad
l_L^{cp}(t,x) =i\gamma^0\gamma^2 ~ l^\star_L(t,-x) ,
\\
u_R^{cp}(t,x)& =i\gamma^0\gamma^2 ~ u^\star_R(t,-x),\qquad\qquad
d_R^{cp}(t,x) =i\gamma^0\gamma^2 ~ d^\star_R(t,-x) ,
\\
\nu_R^{cp}(t,x)& =i\gamma^0\gamma^2 ~ \nu^\star_R(t,-x), \qquad\qquad
e_R^{cp}(t,x) = i\gamma^0\gamma^2 ~ e^\star_R(t,-x) ,
\ea
and
\ba
A^{cp}(t,x) = A^T(t,-x),&\qquad
\phi^{cp}_1(t,x)= \phi^\star_1(t,-x),\qquad
\phi^{cp}_2(t,x) = \phi^\star_2(t,-x),
\ea
In order to reduce statistical noise, we will construct an explicitly CP-symmetric ensemble of initial conditions, where for every randomly generated initial configuration of $\phi_{1,2}$ and $u$, $d$, $e$, $\nu$ and $E_i$, we include its CP-transformed. 

The CP-symmetric ensemble ensures, that when C-symmetry is turned off in the Higgs potential, the ensemble averages of the CP-asymmetric observables $N_W^{1,2}$ and $N_{\rm cs}$ are identically zero. We have checked numerically that this is indeed the case (see also \cite{saffin,mou}). This also means that there is no spurious CP-violation from counter terms or discretisation. 

\section{Results}
\label{sec:results}

We implement the action on a space-time lattice, and derive the classical equations of motion for the bosonic fields and the linear operator equations for the fermions (see appendix A, Eqs.~(\ref{eq:feom}, \ref{eq:geom}, \ref{eq:heom})). We use Wilson fermions on the lattice, to get rid of spatial fermion doublers, while suppressing the time-like doublers through a small time step\footnote{The time-like doubler modes then have very large frequency and stay un-excited if initialised that way.}. The bosonic equations depend on the fermion bilinears, through their quantum expectation values. These are computed from the real-time solutions of the fermion operator equations through the "ensemble" fermion method \cite{borsanyi}. For details on the numerical implementation of this method, we refer to \cite{mou}, and to the appendices in the present paper. 

The fermion expectation values are formally divergent in the continuum limit. This can be resolved by introducing counter terms in the bosonic equations of motion. This is described in Appendix \ref{app:counterterms}.

For each random initial realisation of the Higgs field and fermion ensemble, we solve the evolution equations in temporal gauge $A_0=0$. We use a spatial lattice of size $(L v)^3=(n_x av)^3$ with $N_x=32$ and $av=1.2$ denoting the Higgs vev in lattice units. MPI-parallelised runs each take about 8 hours on 80 cpu's. The total cpu-time used for  400 CP-conjugate pairs is therefore approximately  500.000 hours. 

\begin{figure}
\begin{center}
\includegraphics[width=10cm]{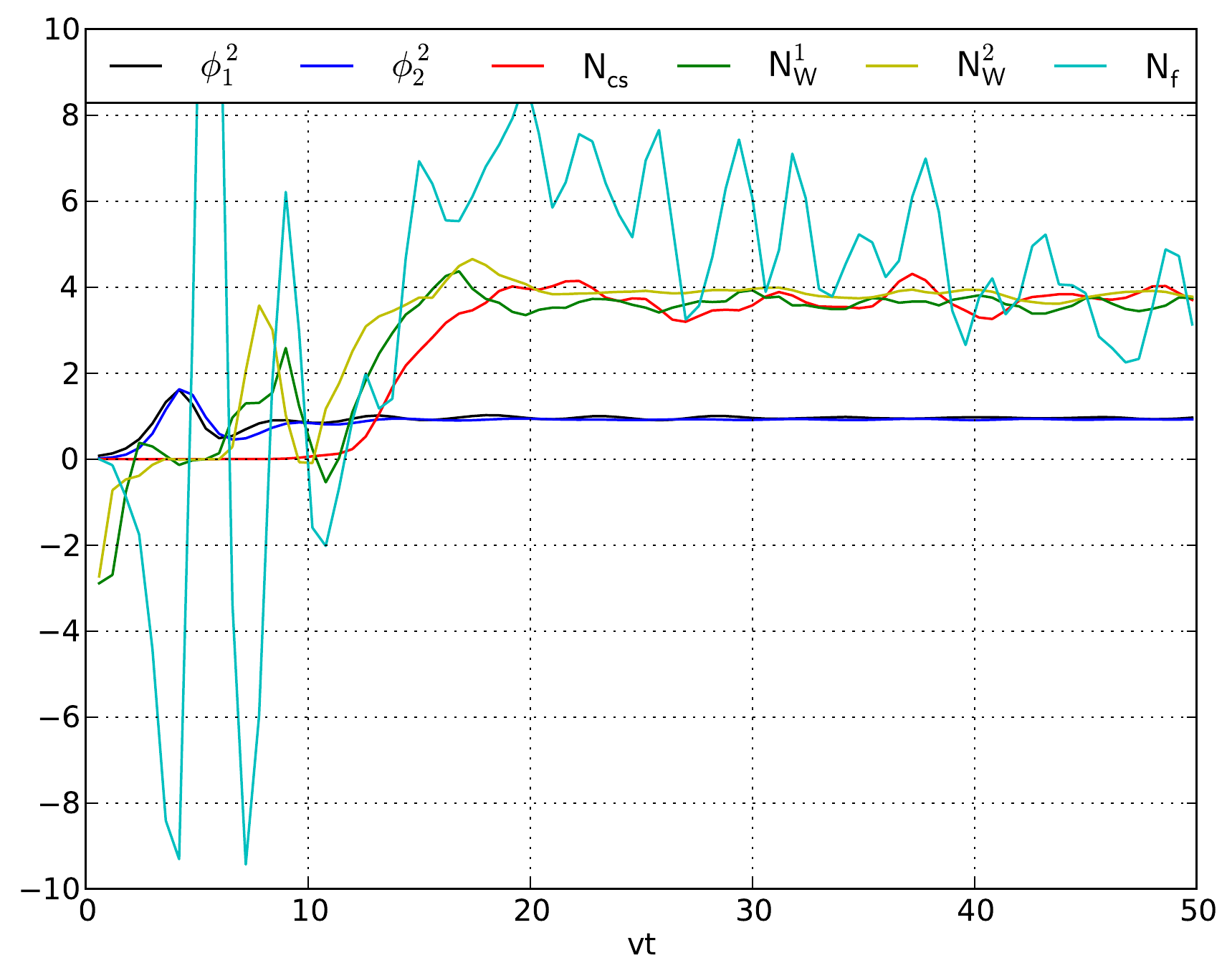}
\caption{Basic observables for a single bosonic realisation; $\phi^2_{1,2}$, $N_W^{1,2}$, $N_{\rm cs}$ and a not very well converged $N_f$.}
\label{fig:NFcomp1}
\end{center}
\end{figure}

\subsection{Which observable to use: $N_W$, $N_{\rm cs}$ or $N_f$?}
\label{sec:obs_choice}

In Figure \ref{fig:NFcomp1}, we show the average Higgs field, the winding numbers, the Chern-Simons number and the fermion number for a single configuration. As was demonstrated in \cite{mou} the fermion number follows the Chern-Simons number in accordance with the anomaly equation, but the statistical noise of the observable $N_f$ is substantial. This noise originates in the UV of the lattice operator, and is largely white noise that may be reduced by increasing the fermion ensemble. What is important for our purposes here is that the back-reaction on the bosonic degrees of freedom has converged, and this is achieved to a reasonable degree for $N_f\simeq 4000$ \cite{mou}.

\begin{figure}
\begin{center}
\includegraphics[width=12cm]{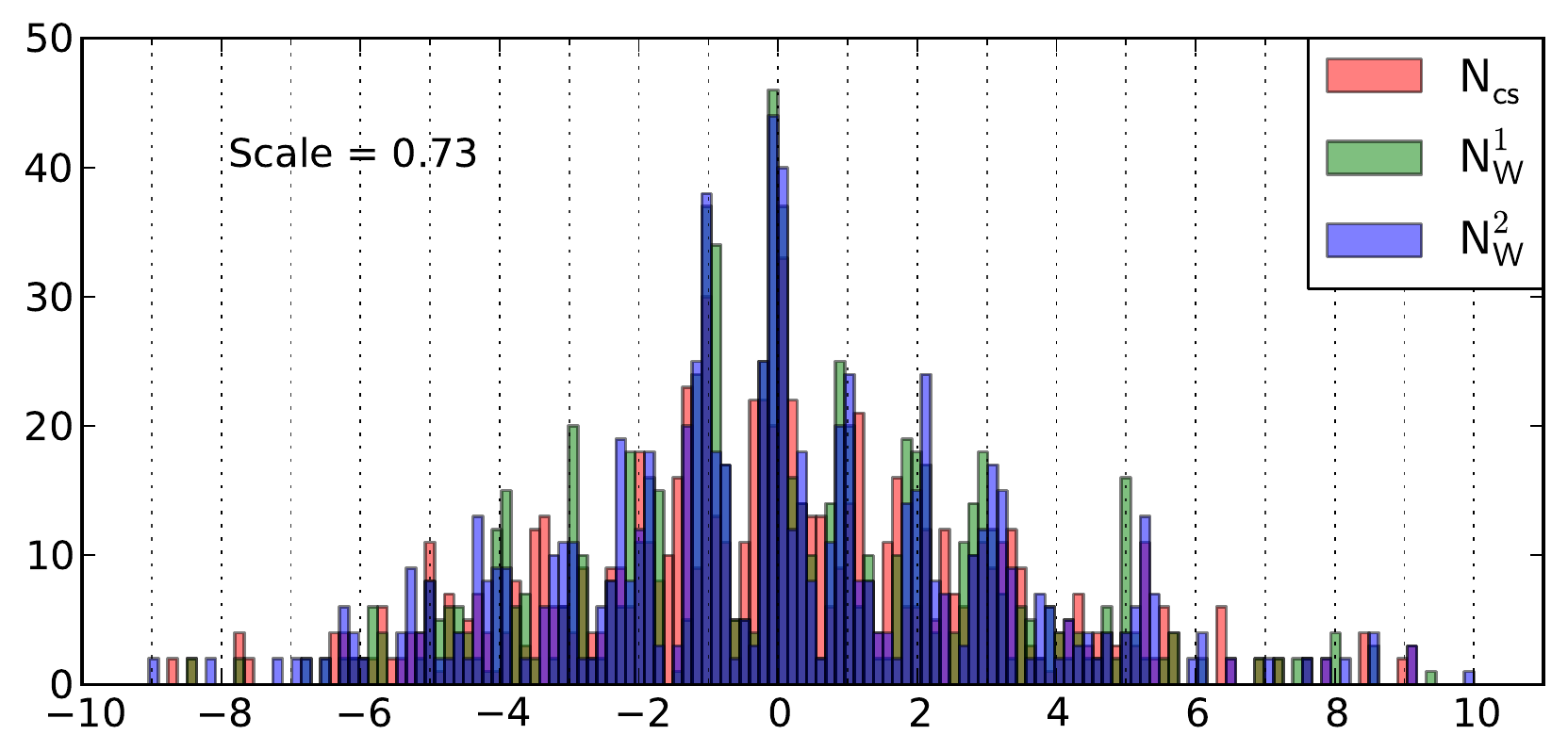}
\caption{Histograms of final values of $N_{\rm cs}$ and $N_W^{1,2}$. The distribution peaks around integers, when normalised by a multiplicative lattice correction factor of $0.73$.}
\label{fig:hist}
\end{center}
\end{figure}

The bosonic fields evolve in a plausible way: The two Higgs fields roll off the potential barrier at $\phi_{1,2}=0$, and oscillate and damp asymptotically to their respective vacuum expectation values, which is here normalised to unity (blue and black). The Higgs transition is over by $vt\simeq 10$, but the topological observables do not settle until $vt\simeq 20$, i.e. after twice as long. Chern-Simons number (red) changes from zero to (in this case) 4 in the interval $10<vt<20$, following the two Higgs winding numbers, but lagging a little behind (yellow and green). Winding number changes most readily when the average Higgs field is small, and we indeed see a failed attempt at a winding number transition around the first Higgs minimum at $vt\simeq 7$.

Both winding numbers and Chern-Simons number suffer some amount of lattice corrections, which tend to reduce the vacuum asymptotic values to somewhat below integer.  
Figure \ref{fig:hist} shows a histogram of final winding and Chern-Simons numbers, over all the configurations simulated. We see a clear set of peaks, corresponding to integer values. In fact the values in the figure have been rescaled by $1/0.73$ to compensate for lattice artefacts. The overall distribution of integers looks roughly Gaussian, and although the difference is not large, the winding numbers tend to be somewhat more peaked than Chern-Simons number for each integer. Because of the statistics noise on the fermion number and the lattice artefacts, we choose to use $N_W^{1,2}$ as the cleanest observables. As we will see below, since we are looking for complete integer flips, this choice is convenient, but $N_{\rm cs}$ could be used as well. 

\subsection{Looking for flips}
\label{sec:flips}

\begin{figure}
\begin{center}
\includegraphics[width=10cm]{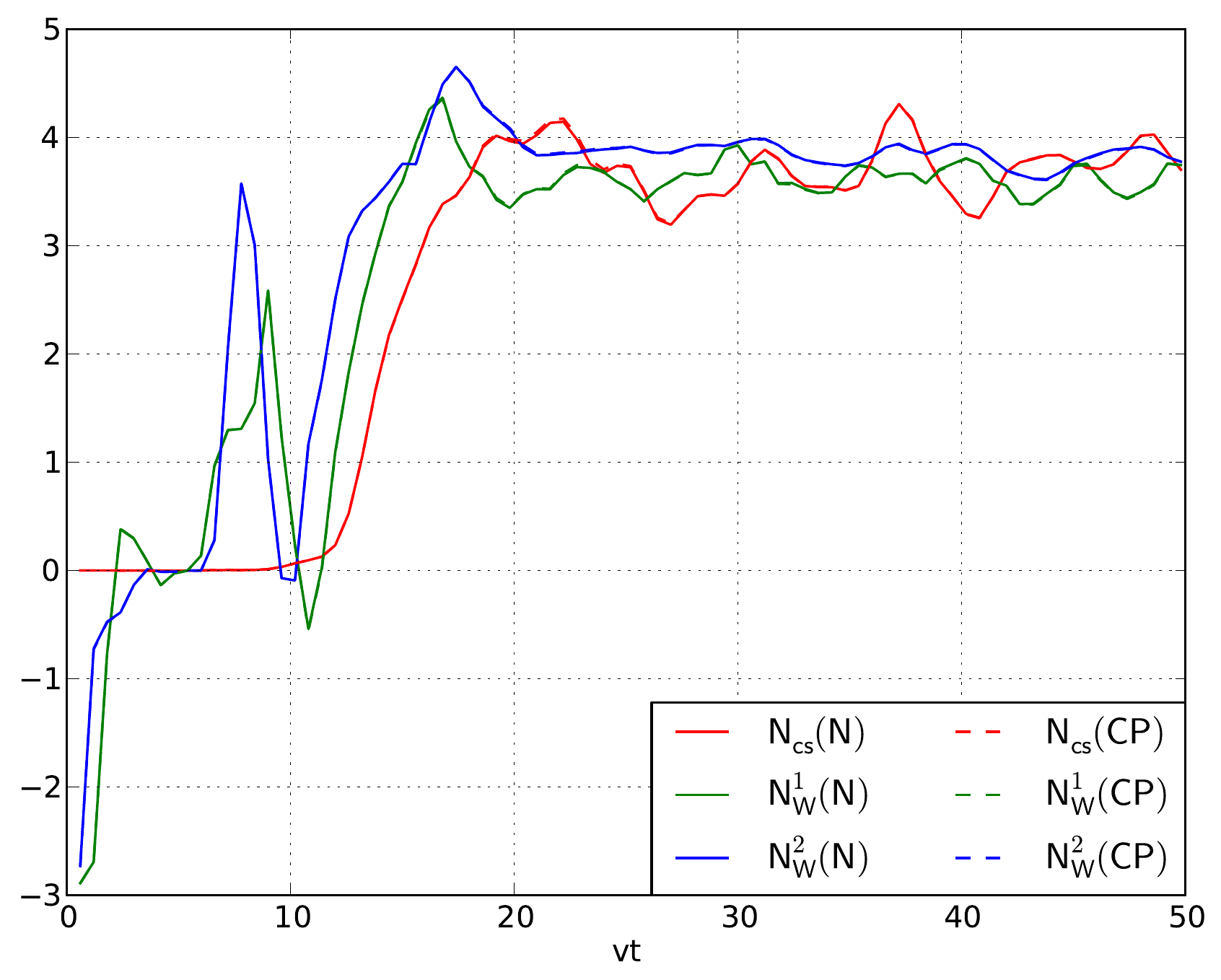}
\caption{A configuration and its CP conjugate, where no net asymmetry is produced. The sign of the observables for the CP conjugate has been flipped.}
\label{fig:noflip}
\end{center}
\end{figure}
Next, we demonstrate that "flips" do occur, configurations where the observables and the CP conjugate configuration observables do not add up to zero after the transition. First, Figure \ref{fig:noflip} shows a pair of configurations that average out to zero in both winding number and Chern-Simons number. The observables for the CP-conjugate pair have been given the opposite sign for comparison. We see that not only is there no flip, the observables are also closely similar, almost to the point that they are indiscernible. The effect of C, P and CP violation stays quite small.

\begin{figure}
\begin{center}
\includegraphics[width=10cm]{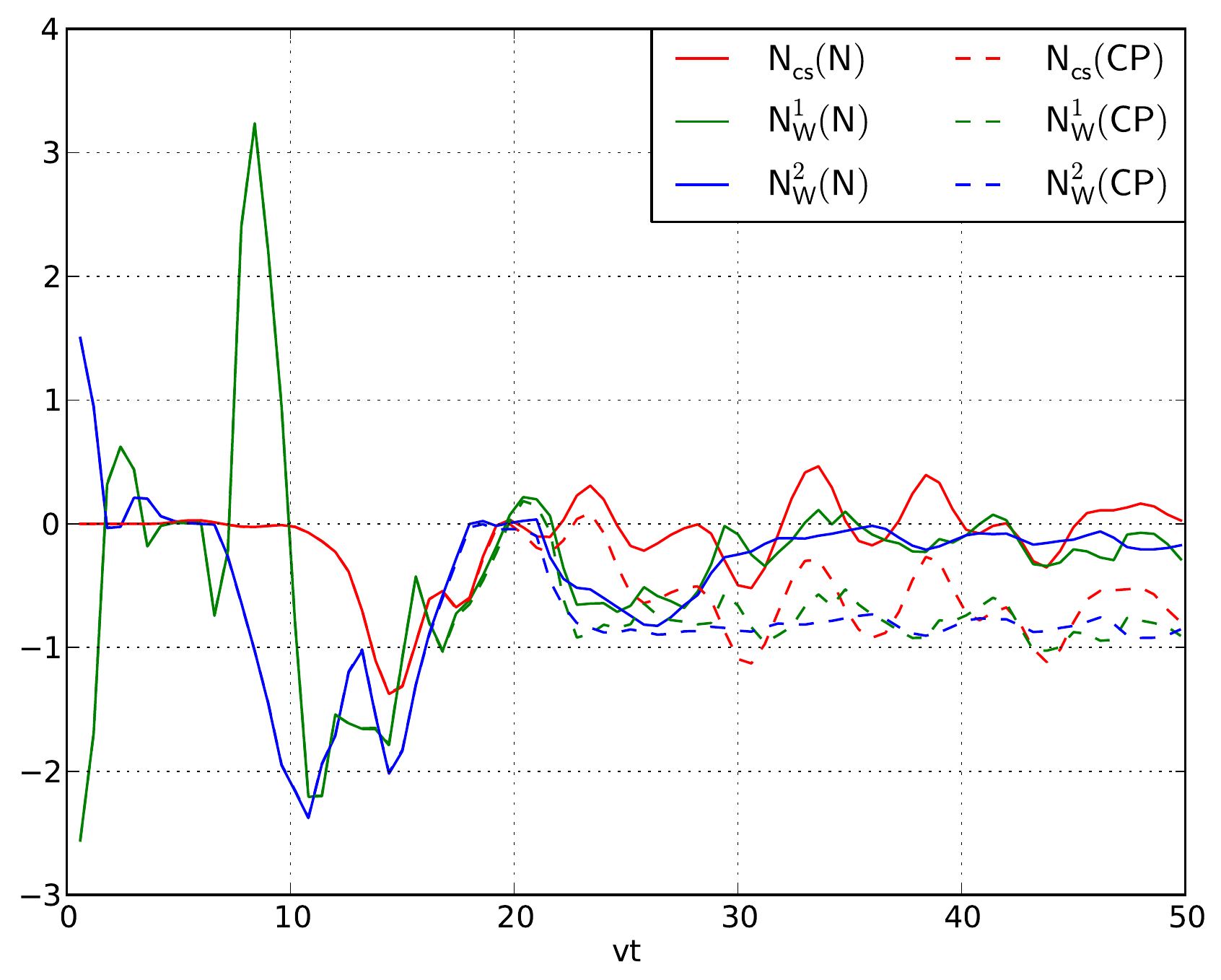}
\caption{A configuration and its CP conjugate, where a flip occurs and an asymmetry is produced. }
\label{fig:plusflip}
\end{center}
\end{figure}

In Figure \ref{fig:plusflip}, we see a case where observables of the pair are again closely similar (up to an overall sign), but suddenly around $vt=30$, the CP-conjugate configuration (dashed lines) stays at $N_W=(-)1$ whereas the original configuration flips back to zero. These are the kind of events we are looking for, and these will stay as a permanent asymmetry in the fermions. In the present case the flip is of $+1$ in the total winding number. Closer inspection shows that the mismatch begins to accumulate around $vt=20$, having been vanishing for earlier times. Also, the transition seems to be controlled by the evolution of $N_{\rm cs}$, which first splits away from $N_W^{1,2}$ in both configurations, but for one the winding number moves to rejoin the Chern-Simons number; in the other Chern-Simons number moves to rejoin the winding number. 

This sort of event is reminiscent of the early work of \cite{turok}. A configuration with mismatched $N_{\rm cs}$ and $N_W$ is gauge equivalent to a $N_{\rm cs}\rightarrow 0$, $N_W\rightarrow N_W-N_{\rm cs}$ localised texture. One may then ask under which circumstances such a texture decays by the winding number going away, rather than the gauge field moving to screen the topological charge, i.e. change Chern-Simons number. In \cite{turok} it was argued that there is a typical physical size of the texture, deciding one or the other, and that this critical size is shifted in the presence of CP-violation. A similar behaviour was observed in \cite{vandermeulen}.

In our setup, C and P violation from the fermions acts on the gauge fields and C violation enters in the scalar dynamics from the potential. We may therefore identify our flips with events where a texture is created at random, and its unwinding is just biased enough by C/P and C violation that it goes opposite ways for the CP-conjugate pair.

\begin{figure}
\begin{center}
\includegraphics[width=14cm]{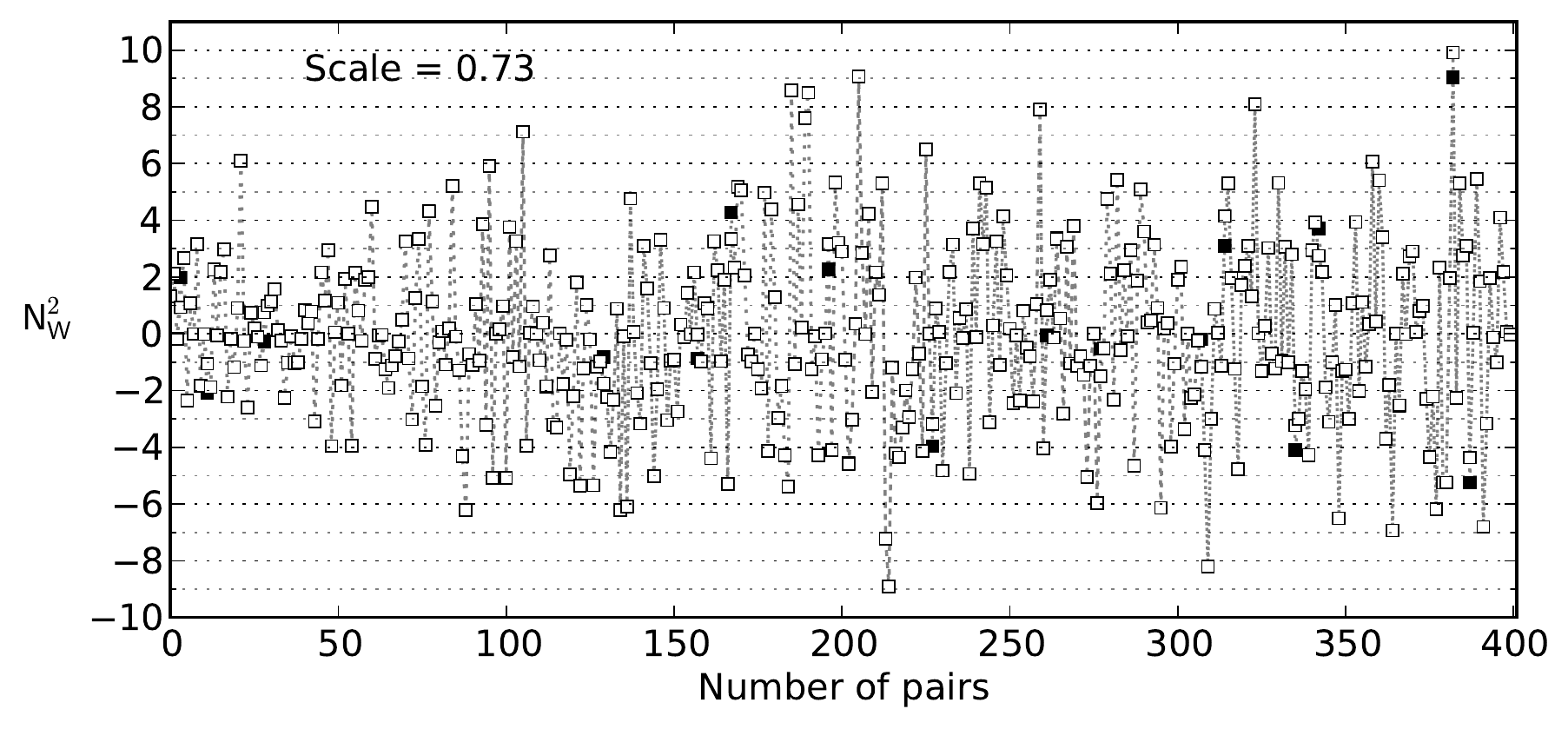}
\caption{Final values of $N_W^2$ for configurations (black) and their CP-conjugate (white).}
\label{fig:allflips}
\end{center}
\end{figure}

In Figure \ref{fig:allflips}, we display the final values of $N_W^2$ for all 400 configurations (black)  and their CP-conjugate (white), rescaled by the lattice artefact correction $1/0.73$. Firstly, we notice how precisely most of them overlap, so that the black symbols are practically hidden. But a few (16) do not match, and they correspond to flips, 6 times $+1$ and 10 times $-1$.

\subsection{Averages}
\label{sec:averages}

\begin{figure}
\begin{center}
\includegraphics[width=14cm]{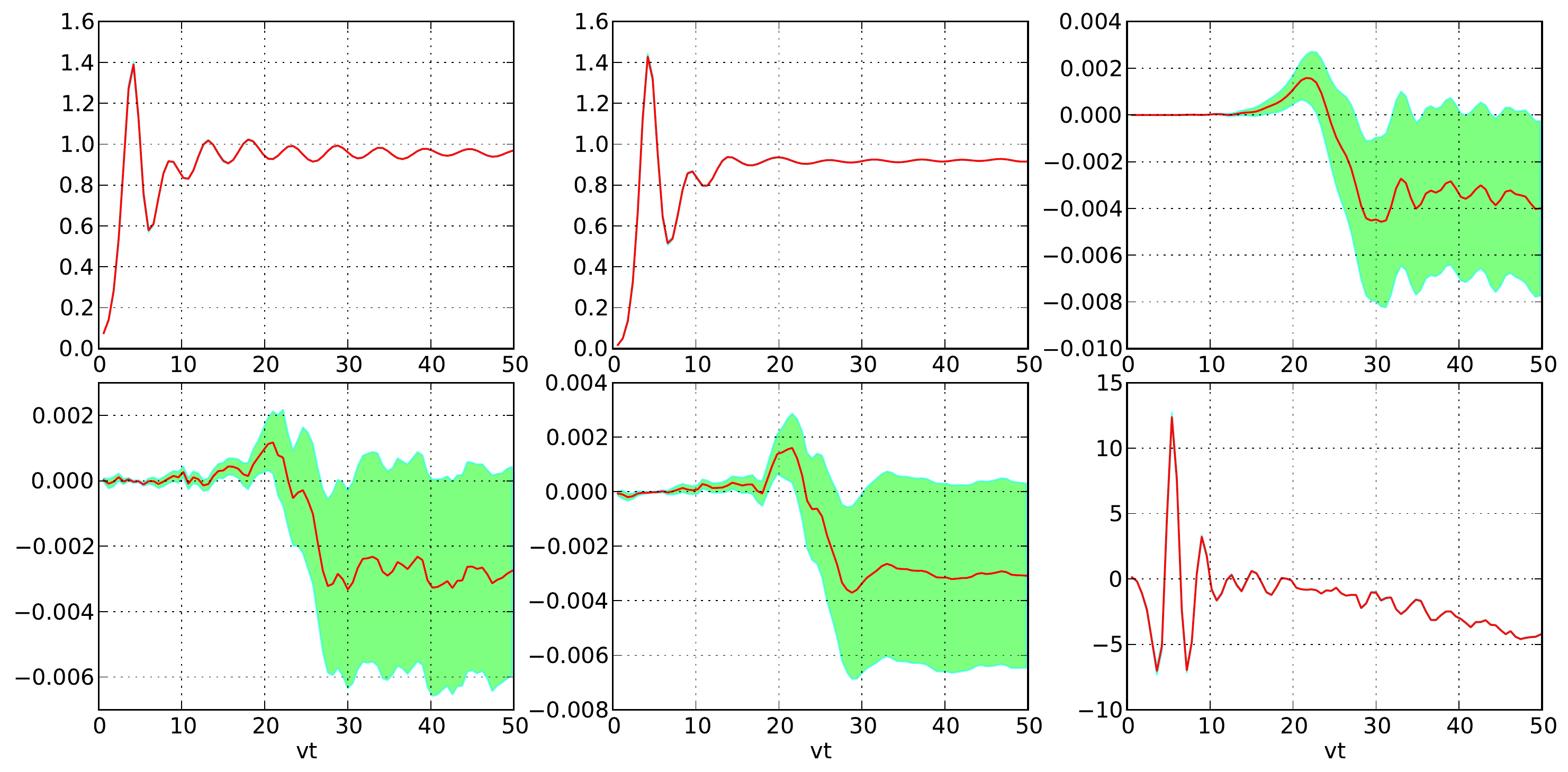}
\caption{Clockwise: The Higgs fields, Chern-Simons number, fermion number and winding numbers, when averaged over the entire bosonic ensemble.}
\label{fig:ave1}
\end{center}
\end{figure}

We conclude by showing in Figure \ref{fig:ave1}  the observables in time, averaged over the entire ensemble of bosonic realisations. The green bands correspond to one standard deviation. 
We see that the average Higgs fields $\phi_{1,2}^2$ are very well behaved statistically, and describe a strongly damped oscillation, approaching the equilibrium expectation value. Because of finite temperature corrections, this is slightly smaller than the vevs, to which the Higgs fields are normalised. Although the Chern-Simons number $N_{\rm cs}$ seems to have a broad statistical uncertainty, the average evolution looks strikingly similar to the purely bosonic simulations \cite{smit1,smit2,smit3,wu1,wu2} (up to an overall sign). First a semi-exponential growth until $vt\simeq 20$, then a dip to a final value of the opposite sign. In \cite{smit1}, the initial rise was explained in terms of a linear treatment of the effective CP-violation, and the general transfer of energy to the gauge field during the spinodal preheating. That this initial behaviour should carry over from using effective bosonic operators to the full fermion dynamics is perhaps not surprising. What is maybe more interesting is that also the qualitative "bouncing" to the opposite side occurs with the dynamical fermions providing the C- and P- violation.  A similar behaviour is observed for the winding numbers. 
Having discovered and asymmetry of  6-10=-4 in 400 pairs, we would expect final winding and Chern-Simons numbers of  $-4/800\simeq-0.005$. This does not follow directly from these averaged observables, without including the lattice artefact correction 0.73, which itself is only approximate. Including this, the integer counting corresponds to $N_W^{1,2}\simeq 0.004$ at the final time. The naive statistical error bars are as large as the signal, and must therefore be taken as inconclusive. We do suspect that the naive standard deviation is an over-estimate of the statistical error, and therefore count these results as suggestive. 

The fermion number is badly behaved, but seems to agree between different bosonic realisations. This shows that the lattice fermion number observable is indeed dominated by UV modes, that are all initialised identically irrespective of the random background bosonic field. 

\subsection{Computing the asymmetry}
\label{sec:BAlattice}

Our simulations are performed on a lattice of physical size $v^3 V=v^3 L^3= n_x^3\times (av)^3$. Given $N_+-N_-$ flips in $N_{\rm tot}$ pairs of configurations, this constitutes a baryon asymmetry of
\ba
n_B=\frac{1}{V}\frac{N_+-N_-}{2N_{\rm tot}}.
\ea
The total energy available in the simulation is $V_0=(226\textrm{ GeV})^4$, which should be distributed over all the active degrees of freedom
\ba
V_0=\frac{\pi^2}{30} g^* T^4 \rightarrow 7.04\times n_\gamma=\frac{2\pi^2}{45}g^*T^3 = 1.01\times (g^*)^{1/4}V_0^{3/4}. 
\ea
Then the baryon to photon ratio is
\ba
\frac{n_B}{n_\gamma} = 6.97 \,\frac{1}{(av)^3n_x^3}\frac{N_+-N_-}{2N_{\rm tot}} (g^*)^{-1/4}\left(\frac{v}{V_0^{1/4}}\right)^3.
\ea
Using $g^*=28$ (in the reduced SM simulated here, $g^*\simeq100$ in the full Standard Model), $v=246$ GeV, $V_0^{1/4}\simeq 226$ GeV on a lattice $n_x=32$, $av=1.2$, we find 
\ba
\eta = 6.9\times 10^{-5}\times
\frac{N_+-N_-}{2N_{\rm tot}}\left(1\pm \sqrt{\frac{N_++N_-}{(N_+-N_-)^2}}
\right),
\ea
where the one standard deviation error applies for $N_{\rm tot}\gg 1$.
The observed asymmetry of $6.0\times 10^{-10}$ corresponds to one uncancelled flip in $10^5$ configurations. This is numerically a hopeless task to find, since only fully completed flips count. An overall shift close to a minimum in the Chern-Simons number potential is not meaningful. We were able to simulate  400 such configuration pairs, so that one uncancelled flip amounts to $\eta\simeq 8.6\times 10^{-8}$. However, the statistical error is large if there are many flips cancelled out ($N_++N_-)>(N_+-N_-)^2$.

\section{Conclusion:}
\label{sec:conclusion}

\subsection{ Can we see a net asymmetry?}

In a ensemble of  400 pairs of realisations, we found  $N_+=6$ flips with +1 difference in $N_W$;  and  $N_-=10$ with -1. We therefore have 
\ba
\eta \simeq -3.5\times 10^{-7}\times \left(1.0\pm 1.0\right),
\ea
about 600 times as big as the observed asymmetry; but also consistent with zero, if these simple statistical error estimates are to be relied upon. Taking the average value as physical, one may readily compare to the results of \cite{wu2}, obtained by replacing the fermion degrees of freedom by an effective C/P breaking bosonic operator of the form
\ba
\delta V = \frac{\delta_{C/P}}{16\pi^2 m_W^2}i(\phi_1^\dagger\phi_2-\phi_2^\dagger\phi_2)\textrm{Tr}F^{\mu\nu}\tilde{F}_{\mu\nu}.
\ea
The expectation is, that to leading order in a gradient expansion, this effective term arises upon integrating out the fermions (see however \cite{salcedo,smit,brauner}). In this way one could in principle compute the coefficient $\delta_{C/P}$. In the present work including fermions, this term and all higher order terms are in principle included at one-loop\footnote{And presumably to higher loop order with only classical bosonic internal lines.}. 

Our values of $(\alpha_1,\alpha_2,\alpha_3)$ correspond exactly to the the lowest left black point in the bottom right panel of Figure 1 in \cite{wu2}.
The simulations in that paper were done at $\mu=\sqrt{\textrm{Im}(\mu_{12}^2)}=100$ GeV (half of the value in the present work). Typical asymmetries quoted in \cite{wu2} are $|\eta|= (1-2)\times 10^{-4}$, with a maximal value of
\ba
\eta=-1.1\times 10^{-5}\times \delta_{C/P}.
\ea
Under the assumption that the precise value of $\mu$ is not decisive, and to the extent that the fermion degrees of freedom can be represented by such a bosonic operator, we conclude that in this scenario, the effective coefficient is of the order of 
\ba
\delta_{C/P}^{\rm eff}\simeq 0.03.
\ea
This is much smaller than one, but not nearly as small as for Standard Model CP-violation at finite temperature \cite{brauner}, which seems to totally rule out electroweak baryogenesis in the Standard Model. This is likely due to the large Yukawa couplings and the fact that CP-violation in the SM kicks in at higher loop order than in the present case of a tree-level CP breaking effect combined with a tree-level C and P breaking effect.
 
 \subsection{Outlook}
 
In conclusion, we have successfully developed a method to numerically compute the baryon asymmetry from first principles in a viable reduced Standard Model. The numerical effort is significant, and with our available numerical resources, we were only barely able to see a signal. But clearly, the effect of CP-violation is present, and using an explicitly CP-symmetric ensemble of realisations allows us to see this. Using a fermion ensemble of $N=4000$ for each bosonic realisation is a cautious choice, where we may be confident that the fermion back-reaction on the bosonic evolution is well converged. This back-reaction is important, since it provides the breaking of P absent in the Higgs potential and necessary for the asymmetry to be generated. 

We tentatively find an asymmetry of $\approx600$ times the observed one, but with a very large statistical uncertainty. An increase in computing time of a factor 10 is necessary to pin this down convincingly. We are currently seeking to acquire such resources for future work. 

Obvious applications and extensions of this method are to consider other experimentally allowed regions of two-Higgs doublet parameter space; including all three families of fermions with the physical values of the Yukawa couplings; including the mechanism responsible for the cold spinodal transition (low-scale inflaton, singlet Higgs field); and the interaction of fermions with a bubble wall in a first order phase transition, relevant for "hot" electroweak baryogenesis. The inclusion of U(1) and SU(3) gauge fields may also be envisaged, but then the transformation from a chiral to a vector theory no longer works, and one will have to in another way circumvent the issue of chiral fermions on the lattice. 
 
\vspace{0.4cm}

\noindent {\bf Acknowledgments:} ZGM would like to thank CSC for financial support. PMS thanks STFC for financial support. AT is supported by the Villum Kann Rasmussen Foundation. The numerical work was performed in part at Dirac facility COSMOS (UK) and the Notur facility Abel (Norway).

\appendix

\section{Lattice implementation}
\label{sec:latticeeom}

Putting chiral fermions on the lattice is notoriously difficult, but because of the pseudo-reality of SU(2), it is possible to bypass this problem by defining (in the continuum) a new set of fermion fields, in the following way:
\ba
\Psi_L&=&q_L,\qquad \Psi_R=\epsilon\mathcal{C}^{-1}\bar l_L^T\qquad\Rightarrow \quad 
          \quad l_L=-\epsilon\mathcal{C}^{-1}\bar\Psi_R^T,
          \quad\bar\l_L=\Psi^T_R\mathcal{C}\epsilon^{-1},\\
\chi_R&=&u_R,\qquad
\chi_L=\mathcal{C}^{-1}\bar e_R^T\qquad\Rightarrow e_R=\mathcal{C}^{-1}\bar \chi_L^T,\quad\bar e_R=-\chi_L^T\mathcal{C},\\
\xi_R&=&d_R,\qquad 
\xi_L=-\mathcal{C}^{-1}\bar \nu_R^T\qquad\Rightarrow \nu_R=-\mathcal{C}^{-1}\bar\xi_L^T,\quad\bar \nu_R=\xi_L^T\mathcal{C},
\\
\Phi &=& \left( \tilde \phi, \phi\right), \qquad  \tilde \phi = \epsilon \phi^\star, \quad \epsilon = i\sigma_2,
\ea
This amounts to a charge conjugation on the left-handed component. Now, instead of having four right-handed singlets and two left-handed doublets, we have a full doublet (left- and right-handed) and two full singlets. This turns the reduced Standard Model into a vector theory rather than a chiral one, and the lattice implementation becomes similar to 2-colour QCD. 

The action now reads:
\ba
\label{eq:action1}
S_G &=& \sum_{x,t}\beta_G^t\sum_i\left(1-\frac{1}{2}\Tr \, U_{0i,x}\right)-\frac{\beta_G^s}{2}\sum_{ij}\left(1-\frac{1}{2}\Tr\,  U_{ij,x}\right),\\
S_H &=&  \sum_{x,t} \sum_{n}\left(
 \frac{\beta_H^t}{2}\Tr\left[(D_0 \Phi_n)^\dagger D_0\Phi_n\right]
-\frac{\beta_H^s}{2}\Tr\left[(D_i \Phi_n)^\dagger D_i\Phi_n\right]
\right)
\nonumber\\ & & \quad
-\beta_RV(\Phi_1,\Phi_2) + \beta_Y C(\Phi_1, \Phi_2) 
,\\
S_F&=&\sum_{x,t}-\left[\bar{\Psi}\gamma^{0}\tilde{D}_{0}\Psi+\bar{\chi}\gamma^{0}\tilde{\partial}_{0}\chi+\bar{\xi}\gamma^{0}\tilde{\partial}_{0}\xi\right]-\frac{a_t}{a}\left[\bar{\Psi}\gamma^{i}\tilde{D}_{i}\Psi+\bar{\chi}\gamma^{i}\tilde{\partial}_{i}\chi+\bar{\xi}\gamma^{i}\tilde{\partial}_{i}\xi\right]\nonumber\\
& &\quad~ +\frac{ra_t}{2a}\, \left[\bar{\Psi}D_i^{'}D_i\Psi+\bar{\chi}\partial_i^{'}\partial_i\chi+\bar{\xi}\partial_i^{'}\partial_i\xi\right]
-\beta_YY(\Psi,\chi,\xi,\Phi_1,\Phi_2),
\label{eq:action2}
\ea
with the 2-Higgs scalar potential
\ba
V(\Phi_1,\Phi_2)
&=&
-\frac{a^2\mu_{11}^2}{4}  \Tr\left[\Phi_1^\dagger\Phi_1\right]  
-\frac{a^2\mu_{22}^2}{4} \Tr\left[\Phi_2^\dagger\Phi_2\right] 
-\frac{ a^2\mu_{12RE}^2 }{2}             \Tr\left[\Phi_1^\dagger\Phi_2\right] 
\nonumber\\ & &
+\frac{ a^2\mu_{12IM}^2}{2}              \Tr\left[\Phi_1^\dagger\Phi_2i\sigma_3\right] 
\nonumber\\ & &
+\frac{\lambda_1}{8}\left(\Tr\left[\Phi_1^\dagger\Phi_1\right] \right)^2 
+\frac{\lambda_2}{8}\left(\Tr\left[\Phi_2^\dagger\Phi_2\right] \right)^2 
\nonumber\\ & &
+\frac{\lambda_3}{4}\Tr\left[\Phi_2^\dagger\Phi_2\right] \Tr\left[\Phi_1^\dagger\Phi_1\right] 
\nonumber\\ & &
+\frac{ \lambda_4}{4}        \left(\Tr\left[\Phi_1^\dagger\Phi_2\right]          \right)^2
+\frac{ \lambda_4}{4}        \left(\Tr\left[\Phi_1^\dagger\Phi_2i\sigma_3\right] \right)^2
\nonumber\\ & &
+\frac{ \lambda_{5RE}}{4}    \left(\Tr\left[\Phi_1^\dagger\Phi_2\right]          \right)^2
-\frac{ \lambda_{5RE}}{4}    \left(\Tr\left[\Phi_1^\dagger\Phi_2i\sigma_3\right] \right)^2
\nonumber\\ & &
-\frac{ \lambda_{5IM}}{2}                \Tr\left[\Phi_1^\dagger\Phi_2i\sigma_3\right] \Tr\left[\Phi_1^\dagger\Phi_2\right] 
,\ea
The Yukawa interactions are now somewhat more complicated, with in general
\ba
\label{eq:yukawa_general}
Y(\Psi, \chi, \xi, \Phi_1, \Phi_2)&=&
~~
 G_1^{u}        \bar\Psi \Phi_1 P_U P_R (\chi , \xi)^T
+G_1^{u\dagger} (\bar\chi , \bar\xi ) P_U\Phi_1^\dagger P_L\Psi
\nonumber\\ & &
+G_2^{u}        \bar\Psi \Phi_2 P_U P_R (\chi , \xi)^T
+G_2^{u\dagger} (\bar\chi , \bar\xi ) P_U\Phi_2^\dagger P_L\Psi
\nonumber\\ & &
+G_1^{d}        \bar\Psi \Phi_1 P_D P_R (\chi , \xi)^T
+G_1^{d\dagger} (\bar\chi , \bar\xi ) P_D\Phi_1^\dagger P_L\Psi 
\nonumber\\ & &
+G_2^{d}        \bar\Psi \Phi_2 P_D P_R (\chi , \xi)^T
+G_2^{d\dagger} (\bar\chi , \bar\xi ) P_D\Phi_2^\dagger P_L\Psi 
\nonumber\\ & &
+G_1^{e}        \bar\Psi \Phi_1 P_U P_L (\chi , \xi)^T
+G_1^{e\dagger} (\bar\chi , \bar\xi ) P_U\Phi_1^\dagger P_R\Psi 
\nonumber\\ & &
+G_2^{e}        \bar\Psi \Phi_2 P_U P_L (\chi , \xi)^T
+G_2^{e\dagger} (\bar\chi , \bar\xi ) P_U\Phi_2^\dagger P_R\Psi 
\nonumber\\ & &
+G_1^{\nu}        \bar\Psi \Phi_1 P_D P_L (\chi , \xi)^T
+G_1^{\nu\dagger} (\bar\chi , \bar\xi ) P_D\Phi_1^\dagger P_R\Psi
\nonumber\\ & &
+G_2^{\nu}        \bar\Psi \Phi_2 P_D P_L (\chi , \xi)^T
+G_2^{\nu\dagger} (\bar\chi , \bar\xi ) P_D\Phi_2^\dagger P_R\Psi
,\ea
in terms of a number of projectors
\ba
P_U=\frac{1+\sigma^3}{2},~~
P_D=\frac{1-\sigma^3}{2},~~
P_R=\frac{1+\gamma^5}{2},~~
P_L=\frac{1-\gamma^5}{2},~~
\ea
For the redefined fields, CP-transformation amounts to
\ba
\Psi^{cp}(t,x) &=& -\gamma^5 i\gamma^0 \gamma^2 \Psi(t,x^p),
\\
\chi^{cp}(t,x) &=&  \gamma^5 i\gamma^0 \gamma^2 \chi(t,x^p),
\\
\xi^{cp}(t,x)  &=&  \gamma^5 i\gamma^0 \gamma^2 \xi(t,x^p),
\\
U_n^{cp}(t,x) &=& U_n^T(t,x^p-n),
\\
\Phi^{cp}(t,x) &=& \Phi^\star(t,x^p),
\ea
The lattice parameters are related to the continuum ones and the lattice spacings as 
\ba
\beta_G^t = \frac{4}{g^2}\frac{a}{a_t}, ~~
\beta_G^s = \frac{4}{g^2}\frac{a_t}{a}, ~~
\beta_H^t = \frac{a}{a_t}, ~~
\beta_H^s = \frac{a_t}{a}, ~~
\beta_R = \frac{a_t}{a}, ~~
\beta_Y = \frac{a_t}{a},
\ea
where $a_t = a dt$ is the lattice spacing in the time direction, with $a$ the spatial spacing. We will refer to $dt$ as the "time step".

It follows that the energy density is given by (for the G(auge), H(iggs) and F(ermion) components, respectively)
\ba
e_G &=& \frac{1}{Va_t}\sum_{x}\beta_G^t\sum_i\left(1-\frac{1}{2}\Tr \, U_{0i,x}\right) + \frac{\beta_G^s}{2}\sum_{ij}\left(1-\frac{1}{2}\Tr\,  U_{ij,x}\right),\\
e_H &=& \frac{1}{Va_t} \sum_{x}\sum_{n}\left(
 \frac{\beta_H^t}{2}\Tr\left[(D_0 \Phi_n)^\dagger D_0\Phi_n\right]
+\frac{\beta_H^s}{2}\Tr\left[(D_i \Phi_n)^\dagger D_i\Phi_n\right]
\right)
+\beta_RV(\Phi_1,\Phi_2)
,\nonumber \\ \\
e_F&=&\frac{1}{Va_t} \sum_{x}\frac{a_t}{a}\left[\bar{\Psi}\gamma^{i}\tilde{D}_{i}\Psi+\bar{\chi}\gamma^{i}\tilde{\partial}_{i}\chi+\bar{\xi}\gamma^{i}\tilde{\partial}_{i}\xi\right]
-\frac{ra_t}{2a}\, \left[\bar{\Psi}D_i^{'}D_i\Psi+\bar{\chi}\partial_i^{'}\partial_i\chi+\bar{\xi}\partial_i^{'}\partial_i\xi\right]
\nonumber \\ & & \quad ~~
+\beta_YY(\Psi,\chi,\xi,\Phi_1,\Phi_2)
 - \beta_Y C(\Phi_1, \Phi_2) 
,\ea
summing to the total energy density.

Because of the field transformation, we compute the baryon and lepton currents of the original theory as the chiral current in the transformed theory
\ba
\left(j^\mu_{(5)}\right)_{\rm C-conjugated}=\left(j^\mu_{(b)}+j^\mu_{(l)}\right)_{\rm Original}&=&i\left[-\bar\Psi\gamma^\mu\gamma^5\Psi+\bar\chi\gamma^\mu\gamma^5\chi+\bar\xi\gamma^\mu\gamma^5\xi\right].
\ea

We now specialise to the Yukawa interactions with a single coupling $y_{uk}$, 
\ba
\label{action:yukawa}
Y(q, l, \phi_1, \phi_2)&=& 
~~ y_{uk}(\bar{q}_L\phi_2 d_R+\bar{l}_L\phi_2 e_R
+ \bar{q}_L\tilde\phi_1 u_R+ \bar{l}_L\tilde\phi_1\nu_R
\nonumber\\&~&
~ ~~~~+\bar d_R\phi_2^\dagger q_L+\bar e_R\phi_2^\dagger l_L
+ \bar u_R\tilde\phi_1^\dagger q_L + \bar\nu_R\tilde\phi_1^\dagger l_L)
\ea
This translates to the lattice theory as
\ba
Y(\Psi, \chi, \xi, \Phi_1, \Phi_2)&=&
y_{uk} 
\left(
 \bar\Psi \Phi_1 P_U P_R (\chi , \xi)^T
+(\bar\chi , \bar\xi ) P_U\Phi_1^\dagger P_L\Psi
\right. 
\nonumber\\ & & \qquad
+\bar\Psi \Phi_2 P_D P_R (\chi , \xi)^T
+(\bar\chi , \bar\xi ) P_D\Phi_2^\dagger P_L\Psi 
\nonumber\\ & & \qquad
+\bar\Psi \Phi_2 P_U P_L (\chi , \xi)^T
+(\bar\chi , \bar\xi ) P_U\Phi_2^\dagger P_R\Psi 
\nonumber\\ & & \qquad
\left.
+\bar\Psi \Phi_1 P_D P_L (\chi , \xi)^T
+(\bar\chi , \bar\xi ) P_D\Phi_1^\dagger P_R\Psi
\right) 
,\ea
Our choice is simply meant as a simplification, and may readily be generalised using the general form (\ref{eq:yukawa_general}). We have carefully considered accidental symmetries of the Yukawa term, to ensure that CP is in fact broken. 

Fermions enter in the bosonic equations of motion as two-point correlators. These are formally divergent, and we introduce counter terms $ct_{11,12}$ in the following way
\ba
\label{action:counterterm}
S_C=C(\Phi_1,\Phi_2)&=&\frac{a^2ct_{11}}{2} \left( \Tr\left[ \Phi_1^\dagger \Phi_1 \right] + \Tr\left[ \Phi_2^\dagger \Phi_2 \right]  \right)
+\frac{a^2ct_{12}}{2} \left( \Tr\left[ \Phi_1^\dagger \Phi_2 \right] + \Tr\left[ \Phi_2^\dagger \Phi_1 \right]  \right) ,\nonumber\\
\ea
We do not consider weaker logarithmic divergences that appear in the gauge equations of motion. $S_C$ should be added to (\ref{eq:action1}-\ref{eq:action2}).

From the complete lattice action, we derive the equations of motion by straightforward variation. For the fermions, linear equations in the bosonic field background:
\ba
\label{eq:feom}
0&=&-\gamma^0\tilde\partial_0\Psi-\frac{a_t}{a}\gamma^i\tilde D_i\Psi+\frac{r a_t}{2a}D_i'D_i\Psi
-y_{uk}\beta_Y  \left[\Phi_1P_U + \Phi_2 P_D\right] P_R (\chi , \xi)^T
\nonumber \\ & & \qquad \qquad \qquad
-y_{uk}\beta_Y  \left[\Phi_2P_U + \Phi_1 P_D\right] P_L (\chi , \xi)^T
\\
0&=&-\gamma^0\tilde\partial_0(\chi,\xi)^T-\frac{a_t}{a}\gamma^i\tilde\partial_i(\chi,\xi)^T+\frac{ra_t}{2a}\partial_i'\partial_i(\chi,\xi)^T
-y_{uk}\beta_Y  \left[P_U\Phi_1^\dagger + P_D\Phi_2^\dagger \right] P_L\Psi 
\nonumber \\ & & \qquad \qquad \qquad
-y_{uk}\beta_Y  \left[P_U\Phi_2^\dagger + P_D\Phi_1^\dagger \right] P_R\Psi 
\ea
For the gauge field
\ba
\label{eq:geom}
0&=&\frac{1}{2}\beta_G^t \left(E_{n}^a(y)-E_{n}^a(y-0\right)-\frac{1}{2}\beta_G^s\sum_m D_m^{ab'}\Tr \left[i\sigma^b U_{y,m}U_{y+m,n}U^\dagger_{y+n,m}U^\dagger_{y,n}\right]\nonumber\\
\nonumber &&-\beta_H^s\Tr\left[i\sigma^a\Phi_{1,y} (U_{y,n}\Phi_{1,y+n})^\dagger\right]-\beta_H^s\Tr\left[i\sigma^a\Phi_{2,y} (U_{y,n}\Phi_{2,y+n})^\dagger\right]
\\ &&-\frac{a_t}{2a}\left[\bar\Psi_y\gamma^n i\sigma^aU_{y,n}\Psi_{y+n}+\bar\Psi_{y+n}\gamma^nU_{y,n}^\dagger i\sigma^a\Psi_y\right],\nonumber\\
&&+\frac{r a_t}{2a}\left[\bar{\Psi}_y i\sigma^a U_{y,n}\Psi_{y+n}-\bar{\Psi}_{y+n}U_{y,n}^\dagger i\sigma^a \Psi_y\right],
\ea
And for the two Higgs fields
\ba
\label{eq:heom}
0 &=& -\beta_H^t \partial_0' \partial_0 \Phi_1 + \beta_H^s \partial_i' \partial_i \Phi_1
\nonumber \\ \qquad
& & +\frac{\beta_R}{2} \left( a^2\mu_{11}^2 -\lambda_1\Tr\left[\Phi_1^\dagger\Phi_1\right] -\lambda_3\Tr\left[\Phi_2^\dagger\Phi_2\right]  \right) \Phi_1
\nonumber \\ \qquad
& & +\frac{\beta_R}{2} \left( a^2\mu_{12re}^2 -(\lambda_4+\lambda_{5re)}\Tr\left[\Phi_1^\dagger\Phi_2\right] +\lambda_{5im}\Tr\left[\Phi_1^\dagger\Phi_2i\sigma_3\right]  \right) \Phi_2
\nonumber \\ \qquad
& & +\frac{\beta_R}{2} \left( -a^2\mu_{12im}^2 +(-\lambda_4+\lambda_{5re)}\Tr\left[\Phi_1^\dagger\Phi_2i\sigma_3\right] +\lambda_{5im}\Tr\left[\Phi_1^\dagger\Phi_2\right]  \right) \Phi_2i\sigma_3
\nonumber \\ \qquad
& &
-\frac{y_{uk}\beta_Y}{2}k_a \left[ 
 \bar\Psi  k^a P_U P_R (\chi , \xi)^T
+(\bar\chi , \bar\xi ) P_U k^{a\dagger} P_L\Psi 
 \right]
\nonumber \\ \qquad
& &
-\frac{y_{uk}\beta_Y}{2}k_a \left[ 
 \bar\Psi  k^a P_D P_L (\chi , \xi)^T
+(\bar\chi , \bar\xi ) P_D k^{a\dagger} P_R\Psi 
 \right]
\nonumber \\ \qquad
& &
+\beta_Ya^2ct_{11}\Phi_1 + \beta_Ya^2ct_{12} \Phi_2
\\
0 &=& -\beta_H^t \partial_0' \partial_0 \Phi_2 + \beta_H^s \partial_i' \partial_i \Phi_2
\nonumber \\ \qquad
& & +\frac{\beta_R}{2} \left( a^2\mu_{22}^2 -\lambda_2\Tr\left[\Phi_2^\dagger\Phi_2\right] -\lambda_3\Tr\left[\Phi_1^\dagger\Phi_1\right]  \right) \Phi_2
\nonumber \\ \qquad
& & +\frac{\beta_R}{2} \left( a^2\mu_{12re}^2 -(\lambda_4+\lambda_{5re)}\Tr\left[\Phi_1^\dagger\Phi_2\right] +\lambda_{5im}\Tr\left[\Phi_1^\dagger\Phi_2i\sigma_3\right]  \right) \Phi_1
\nonumber \\ \qquad
& & +\frac{\beta_R}{2} \left( a^2\mu_{12im}^2 +(\lambda_4-\lambda_{5re)}\Tr\left[\Phi_1^\dagger\Phi_2i\sigma_3\right] -\lambda_{5im}\Tr\left[\Phi_1^\dagger\Phi_2\right]  \right) \Phi_1i\sigma_3
\nonumber \\ \qquad
& &
-\frac{y_{uk}\beta_Y}{2}k_a \left[ 
 \bar\Psi  k^a P_D P_R (\chi , \xi)^T
+(\bar\chi , \bar\xi ) P_D k^{a\dagger} P_L\Psi 
 \right]
\nonumber \\ \qquad
& &
-\frac{y_{uk}\beta_Y}{2}k_a \left[ 
 \bar\Psi  k^a P_U P_L (\chi , \xi)^T
+(\bar\chi , \bar\xi ) P_U k^{a\dagger} P_R\Psi 
 \right]
\nonumber \\ \qquad
& &
+\beta_Ya^2ct_{11}\Phi_2 + \beta_Ya^2ct_{12}\Phi_1
\ea
In the bosonic equations of motion, all fermion bilinears are replaced by quantum expectation values over the fermion ensemble fields,. 

\section{Counterterms and Renormalization}
\label{app:counterterms}

The counterterms are chosen so as to compensate the backreaction terms of the fermion fields in vacuum, as these terms appear in the bosonic potentials and kinetic terms.   
Appearing as relevant or marginal operators, these counterterms include terms quadratically or logarithmically divergent in the continuum limit. Since we do not intend to take this limit in our simulations, the counter terms simply provide a subtraction of a finite (yet generically large) number. In practice, the most severe (quadratic) divergences are our main concern \ref{action:counterterm}.

When the Higgs fields are constant, as in the vacuum, the integration over the fermion modes yields  
\ba
& &i\int\frac{d^4p}{(2\pi)^4}\ln 
\textrm{Det}\left[ \barray{cc} i\gamma^\mu p_\mu + m_D & y_{uk}K \\ -y_{uk}\gamma^0 K^\dagger \gamma^0 & i\gamma^\mu p_\mu +m_S \earray
\right]_{16\times16}
\nonumber \\
&=& \frac{ct_{11}}{2} \left( \Tr\left[ \Phi_1^\dagger \Phi_1 \right] + \Tr\left[ \Phi_2^\dagger \Phi_2 \right]  \right)
+\frac{ct_{12}}{2} \left( \Tr\left[ \Phi_1^\dagger \Phi_2 \right] + \Tr\left[ \Phi_2^\dagger \Phi_1 \right]  \right)
+ ...,
\ea
where $m_D$ and $m_S$ are the masses of the doublets and singlets respectively, and $K$ denotes the matrix 
\ba
K=\Phi_1P_U\otimes P_R + \Phi_2 P_D\otimes P_R + \Phi_2P_U\otimes P_L + \Phi_1 P_D\otimes P_L
.
\ea
Notice in the backreaction of fermion fields, there is a symmetry between $\Phi_1$ and $\Phi_2$, owing to the fact that $K$ is invariant under the exchange of $\Phi_1$ and $\Phi_2$, and up and down simultaneously, while the latter exchange does not affect the determinant of the matrix. 
Therefore, the simplicity of the counterterms \ref{action:counterterm} can be attributed to the symmetry in Yukawa interactions \ref{action:yukawa}. 

To obtain the expression of $ct_{11}$ and $ct_{12}$, we consider two cases:
\noindent
(a) $\Phi_2=\Phi_1=\frac{\phi}{y_{uk}} I$,
\ba
i\int\frac{d^4p}{(2\pi)^4}\ln 
\textrm{Det}\left[ \barray{cc} i\gamma^\mu p_\mu + m_D & \phi \\  \phi & i\gamma^\mu p_\mu +m_S \earray
\right]_{16\times16}
= 2\frac{ct_{11}}{y_{uk}^2} \phi^2 
+ 2\frac{ct_{12}}{y_{uk}^2} \phi^2
+ ...
.
\ea
The series on the right-hand side contains a constant term which is the result of the free massless fermion, as one can see by simply setting $\phi=0$ on the left-hand side.
Thus, instead of implementing the equation above, we consider its derivative where the constant term is removed,
\ba
i\int\frac{d^4p}{(2\pi)^4}\frac{d}{d\phi}\ln 
\textrm{Det}\left[ \barray{cc} i\gamma^\mu p_\mu + m_D & \phi \\  \phi & i\gamma^\mu p_\mu +m_S \earray
\right]_{16\times16}
= 4\frac{ct_{11}}{y_{uk}^2} \phi 
+ 4\frac{ct_{12}}{y_{uk}^2} \phi
+ ...,
\ea
which leads to
\ba
\label{resa}
& &
  \frac{ct_{11}}{y_{uk}^2} \phi 
+ \frac{ct_{12}}{y_{uk}^2} \phi
+ ...
\nonumber \\
&=&
\int\frac{d^3p}{(2\pi)^3}i 
\int \frac{dp_0}{2\pi}\left[
\frac{-\frac{d\omega_{1a}}{d\phi}}{p_0-\omega_{1a}+i\epsilon}
+\frac{\frac{d\omega_{1a}}{d\phi}}{p_0+\omega_{1a}-i\epsilon}
+\frac{-\frac{d\omega_{2a}}{d\phi}}{p_0-\omega_{2a}+i\epsilon}
+\frac{\frac{d\omega_{2a}}{d\phi}}{p_0+\omega_{2a}-i\epsilon}
 \right] 
\nonumber \\
&=&
-\int\frac{d^3p}{(2\pi)^3} 
\left[
\frac{d\omega_{1a}}{d\phi} +\frac{d\omega_{2a}}{d\phi}
 \right]
.
\ea
The determinant in the equation has been calculated in the manner that
\ba
& &\textrm{Det}\left[ \barray{cc} i\gamma^\mu p_\mu + m_D & \phi \\  \phi & i\gamma^\mu p_\mu +m_S \earray
\right]_{16\times16}
=
\textrm{Det}\left[ p_0+H\right]
=
(p_0^2-\omega_{1a}^2+i\epsilon)^4(p_0^2-\omega_{2a}^2+i\epsilon)^4
,
\nonumber \\
\ea
where $H$ refers to the Hamiltonian 
\ba
H=\left[ \barray{cc} i\gamma^0 i\gamma p + i\gamma^0m_D & i\gamma^0\phi \\ i\gamma^0\phi & i\gamma^0i\gamma p +i\gamma^0m_S \earray
\right]_{16\times 16} 
,
\ea
and its eigenenergies  $\omega_{1a}$ and $\omega_{2a}$ satisfy
\footnote{It is easier to achieve the eigenenergies from $H^2$ than from $H$.}
\ba
\omega_{1a}^2 = p^2 + \phi^2 + \frac{(m_D^2+m_S^2)+\sqrt{(m_D^2-m_S^2)^2+4\phi^2(m_D+m_S)^2}}{2}
,
\\
\omega_{1a}^2 = p^2 + \phi^2 + \frac{(m_D^2+m_S^2)-\sqrt{(m_D^2-m_S^2)^2+4\phi^2(m_D+m_S)^2}}{2}
.
\ea

\noindent
(b) $-\Phi_2=\Phi_1=\frac{\phi}{y_{uk}} I$,
\ba
i\int\frac{d^4p}{(2\pi)^4}\ln 
\textrm{Det}\left[ \barray{cc} i\gamma^\mu  p_\mu + m_D & \phi\sigma^3\otimes \gamma^5\\ -\phi\sigma^3\otimes\gamma^5 & i\gamma^\mu  p_\mu +m_S \earray
\right]_{16\times16}
= 2\frac{ct_{11}}{y_{uk}^2} \phi^2 
- 2\frac{ct_{12}}{y_{uk}^2} \phi^2
+ ...
.
\ea
Similarly, the result is
\ba
\label{resb}
& &
  \frac{ct_{11}}{y_{uk}^2} \phi 
- \frac{ct_{12}}{y_{uk}^2} \phi
+ ...
\nonumber \\
&=&
\int\frac{d^3p}{(2\pi)^3}i 
\int \frac{dp_0}{2\pi}\left[
\frac{-\frac{d\omega_{1b}}{d\phi}}{p_0-\omega_{1b}+i\epsilon}
+\frac{\frac{d\omega_{1b}}{d\phi}}{p_0+\omega_{1b}-i\epsilon}
+\frac{-\frac{d\omega_{2b}}{d\phi}}{p_0-\omega_{2b}+i\epsilon}
+\frac{\frac{d\omega_{2b}}{d\phi}}{p_0+\omega_{2b}-i\epsilon}
 \right] 
\nonumber \\
&=&
-\int\frac{d^3p}{(2\pi)^3} 
\left[
\frac{d\omega_{1b}}{d\phi} +\frac{d\omega_{2b}}{d\phi}
 \right]
,
\ea
where
\ba
\omega_{1b}^2 = p^2 + \phi^2 + \frac{(m_D^2+m_S^2)+\sqrt{(m_D-m_S)^2[(m_D+m_S)^2+4\phi^2]}}{2}
,
\\
\omega_{2b}^2 = p^2 + \phi^2 + \frac{(m_D^2+m_S^2)-\sqrt{(m_D-m_S)^2[(m_D+m_S)^2+4\phi^2]}}{2}
.
\ea
To any order of constant Higgs fields, \ref{resa} and \ref{resb} must be valid.
For our purpose, we only need the leading terms, which read
\ba
 -\int\frac{d^3p}{(2\pi)^3}\phi \left[ \frac{1}{\omega_1} + \frac{1}{\omega_2}  - \frac{(m_D+m_S)^2}{\omega_1\omega_2(\omega_1+\omega_2)}\right]
=\frac{ct_{11}}{y_{uk}^2}\phi
+\frac{ct_{11}}{y_{uk}^2}\phi
,
\\
-\int\frac{d^3p}{(2\pi)^3}\phi \left[ \frac{1}{\omega_1} + \frac{1}{\omega_2}  - \frac{(m_D-m_S)^2}{\omega_1\omega_2(\omega_1+\omega_2)}\right]
=\frac{ct_{11}}{y_{uk}^2}\phi
-\frac{ct_{11}}{y_{uk}^2}\phi
,
\ea
for which we find
\ba
ct_{11} &=& - y_{uk}^2\int\frac{d^3p}{(2\pi)^3} \left[ \frac{1}{\omega_1} + \frac{1}{\omega_2}  - \frac{m_D^2+m_S^2}{\omega_1\omega_2(\omega_1+\omega_2)}\right]
,
\\
ct_{12} &=& ~~  y_{uk}^2\int\frac{d^3p}{(2\pi)^3}\left[  \frac{2m_Dm_S}{\omega_1\omega_2(\omega_1+\omega_2)}\right]
,
\ea
where
\ba
\omega_{1}^2 &=& p^2 + m_D^2
,
\\
\omega_{2}^2 &=& p^2 + m_S^2
.
\ea
Considering the continuum theory, one may immediately notice that $ct_{11}$ is quadratically divergent, as expected; while the divergence of $ct_{12}$ is logarithmic, or null for massless fermions.
Conveniently, the solutions may be applied directly to Wilson fermion, in which case 
$m_D = \frac{1}{2}a r_D (p_1^2+p_2^2+p_3^2)$ and
$m_S = \frac{1}{2}a r_S (p_1^2+p_2^2+p_3^2)$,
with Wilson parameters $r_D$ and $r_S$ for doublets and singlets separately.

\end{document}